\documentclass[twocolumn,showpacs,prb,superscriptaddress]{revtex4}

\usepackage{color}
\usepackage{tabularx}
\usepackage{epsfig}
\usepackage{amsmath}
\usepackage{amssymb}
\usepackage{graphicx}

\begin{document}

\title{Extended scaling for ferromagnetic Ising models with
zero-temperature transitions}

\author{Helmut G.~Katzgraber}
\affiliation {Theoretische Physik, ETH Zurich, CH-8093 Zurich,
Switzerland}

\author{I.~A.~Campbell}
\affiliation{Laboratoire des Collo\"ides, Verres et Nanomat\'eriaux,
Universit\'e Montpellier II, 34095 Montpellier, France}

\author{A.~K.~Hartmann}
\affiliation{Institut f\"ur Physik, Carl-von-Ossietzky-Universtit\"at,
D-26111 Oldenburg, Germany}

\begin{abstract}

We study the second-moment correlation length and the
reduced susceptibility of two ferromagnetic Ising models with
zero-temperature ordering. By introducing a scaling variable motivated
by high-temperature series expansions, we are able to scale data for
the one-dimensional Ising ferromagnet rigorously over the entire
temperature range.  Analogous scaling expressions are then applied to
the two-dimensional fully frustrated Villain model where excellent
finite-size scaling over the entire temperature range is achieved.
Thus we broaden the applicability of the extended scaling method to
Ising systems having a zero-temperature critical point.

\end{abstract}


\maketitle

\section{Introduction}
\label{sec:intro}

Studying the critical behavior of systems that order at
zero temperature is challenging because the typically used
Monte Carlo methods are generally unable to probe the critical
behavior close enough to the zero-temperature critical point for
traditional\cite{yeomans:92} finite-size scaling approaches to
yield precise critical parameters.  It is thus necessary to either
incorporate scaling corrections\cite{beach:05} or find better
approaches to scale the data.\cite{caracciolo:95,ballesteros:99}

Using the intuition gained from high-temperature series expansions
(HTSEs) a scaling approach has been introduced with the aim
of extending the validity of critical scaling expressions to
temperatures well above the critical region.  So far, the approach
has been applied to a number of model systems having finite ordering
temperatures.\cite{campbell:06,campbell:07,campbell:08,berche:08}
Inherently, this approach is ideal to study systems which order at
zero temperature---such as a magnetic system {\em below} the
lower critical dimension---since there only temperatures {\em above}
the critical point can be accessed numerically.  Thus, an important
first step in adapting the extended scaling approach to these systems
involves the appropriate choice of a scaling variable.

In this paper we derive extended scaling relations for two sample
ferromagnetic models with no disorder in the interactions and which
only order at zero temperature.  First, we study the exactly solvable
one-dimensional (1D) Ising ferromagnet and then use insights obtained
to analyze the nontrivial two-dimensional (2D) fully frustrated
Villain Ising model.\cite{villain:77} Data generated using Monte
Carlo simulations for large system sizes and very low temperatures
validate our scaling approach.

The paper is structured as follows. In Sec.~\ref{sec:scale} we discuss
the extended scaling approach and how to adapt it to systems ordering
at zero temperature, illustrating the results with the one-dimensional
case in Sec.~\ref{sec:1d}. In Sec.~\ref{sec:2d} we introduce the
(Villain) fully frustrated Ising model and present details of our
numerical calculations. The numerical data are then analyzed with
the extended scaling approach, followed by concluding remarks.

\section{Extended scaling}
\label{sec:scale}

Conventionally, at a continuous phase transition the power-law critical
behavior of any observable ${\mathcal O}$ in the thermodynamic limit
can be written as\cite{yeomans:92}
\begin{equation}
{\mathcal O}[t(T)] \sim t^{-y}
\label{eq:crit}
\end{equation}
with $t = (T-T_c)/T_c$ as a scaling variable, $T$ the temperature
and $T_c$ the critical temperature at which the phase transition
occurs. The exponent $y$ describes the ``strength'' of the divergence
at $T_c$.  Alternatively, other critical variables which yield the
same limiting behavior at criticality such as 
\begin{equation}
\tau = (T-T_c)/T = [1-\beta/\beta_c] ,
\label{eq:tau}
\end{equation}
where $\beta=1/T$, can be used.  For certain models further scaling
variables have been introduced, e.g.,
\begin{equation}
\tau_{\rm s}= \frac{1}{2}\left[\sinh^{-1}(2\beta)-\sinh(2\beta)\right]
\end{equation}
for the 2D Ising
ferromagnet\cite{comment:escale,orrick:01,caselle:02} or
\begin{equation}
\tau_{\rm g} = [1-(\beta/\beta_c)^2]
\end{equation} 
for spin glasses with zero-mean symmetric interaction distributions.
\cite{daboul:04,campbell:06} All these scaling variables are
proportional to $(T-T_c)$ for $T \to T_c$.  However, $T$ is not
a sensible scaling variable for a zero-temperature ($T_c = 0$)
transition.\cite{baxter:82} Cardy {\em et al.}\cite{cardy:01} gave
a renormalization-group technique rule for low-temperature
limiting scaling variables at zero-temperature transitions.
For example, for the Potts model studied by Cardy {\em et al.} the
appropriate renormalization-group ``temperature'' scaling variable is
proportional to $\exp(-\beta)$.  By analogy, for a 1D ferromagnetic
Ising model for which $T_c = 0$ the scaling variable should be
proportional to $\exp(-2\beta)$.\cite{baxter:82}

Scaling expressions may also include temperature-dependent
prefactors, which are noncritical but can be relevant in
analyses that include a range of temperatures far above (or below)
$T_c$.  For instance, the reduced susceptibility $\chi(\beta)$
measured in numerical simulations (see below) is related to the
thermodynamic susceptibility  $\chi_{\rm th}(\beta)$ (which is the
physically measurable observable) through $\chi(\beta) = \chi_{\rm
th}(\beta)/\beta$.  Thus, critical behavior of the form $\chi(\beta)
\sim t^{-\gamma}$ implies $\chi_{\rm th}(\beta) \sim \beta t^{-\gamma}$,
i.e., with a prefactor $\beta$.

The two thermodynamic limit observables that we discuss here are the
ferromagnetic reduced susceptibility $\chi$ and the second-moment
correlation length $\xi$. The ferromagnetic reduced susceptibility
is given by
\begin{equation}
\chi = N\langle m^2 \rangle ,
\label{eq:chi}
\end{equation}
where
\begin{equation}
m = \frac{1}{N}\sum_{i = 1}^{N}S_i
\label{eq:mag}
\end{equation}
is the magnetization per spin, $N$ is the number of spins in the
system and $\langle \cdots \rangle$ represents a thermal average. The
second-moment correlation length is given by
\begin{equation}
\xi = \left[\frac{\mu_2}{z\chi}\right]^{1/2} ,
\label{eq:xi}
\end{equation}
where
\begin{equation}
\mu_2 = \sum_{i,j = 1}^N
		r_{ij}^{2}
		\langle
		S_{i}S_{j}
		\rangle
\end{equation}
is the second moment of the correlation function, $z$ is the number
of nearest neighbors and $r_{ij}$ is the distance between spins $i$
and $j$.\cite{butera:02} For a hypercubic lattice $z = 2D$, where
$D$ is the space dimension. Note that numerically we measure the
finite-size correlation length [which is equivalent to the expression
presented in Eq.~(\ref{eq:xi})] as
\begin{equation}
\xi = \frac{1}{2 \sin (|{\bf k}_\mathrm{m}|/2)}
\left[\frac{\chi}{\chi({\bf k}_\mathrm{m})}
- 1 \right]^{1/2} ,
\label{eq:xiL}
\end{equation}
where ${\bf k}_\mathrm{m} = (2\pi/L,0)$ is the smallest nonzero wave
vector (here in 2D), and $\chi({\bf k})$ is the  wave-vector-dependent
reduced susceptibility:
\begin{equation}
\chi({\bf k}) = \frac{1}{N} \sum_{i,j = 1}^N \langle S_i S_j
\rangle e^{i {\bf k}\cdot{\bf r}_{ij} } .
\label{eq:chik}
\end{equation}
HTSEs of the Ising
ferromagnet in large space dimensions (i.e., in the mean-field
regime)\cite{campbell:08} show that simple relations for the reduced
susceptibility, namely
\begin{equation}
\chi(\beta) = \tau^{-1} ,
\label{eq:chiMF}
\end{equation}
and for the second-moment correlation length defined in
Eq.~(\ref{eq:xi})
\begin{equation}
\xi(\beta) = \beta^{1/2}\tau^{-1/2}
\label{eq:xiMF}
\end{equation}
are exact for all $T > T_c = 1$. Thus, in this limit with the scaling
variable $\tau$ [Eq.~(\ref{eq:tau})],
the critical power laws for the reduced observables
$\chi_{\rm th}(\beta)/\beta$ and $\xi(\beta)/\beta^{1/2}$ hold exactly
over the entire range of $\beta$ from $\beta_c$ to zero. In
finite-dimensional ferromagnetic systems---if the same basic variables
and expressions are used\cite{campbell:06,campbell:07,campbell:08}
with the modification necessary to give the right high-temperature
limits---one obtains ``extended scaling'' equations in which the
leading terms are
\begin{equation}
\chi(\beta) = C_{\chi}\tau^{-\gamma} +(1-C_{\chi})
\label{eq:chiext}
\end{equation}
and
\begin{equation}
\xi(\beta) = \beta^{1/2}[C_{\xi}\tau^{-\nu} + (1-C_{\xi})] \; .
\label{eq:xiext}
\end{equation}
In Eqs.~(\ref{eq:chiext}) and (\ref{eq:xiext}) $C_{\chi}$ and $C_{\xi}$
are critical amplitudes and $\gamma$ and $\nu$ are the standard
critical exponents.\cite{yeomans:92} With the appropriate critical
parameters (critical temperature, critical exponents, as well as
critical amplitudes) these expressions are exact {\em by construction}
at the $\beta \to \beta_c$ and $\beta \to 0$ limits. Elsewhere,
the expressions are not exact but have been shown to give good
approximations to the true behavior for the entire paramagnetic
temperature region. By introducing small correction terms these
approximations can be improved considerably.

The expression for finite $\beta_c$ given in Eqs.~(\ref{eq:chiext})
and (\ref{eq:xiext}) cannot be used for systems with $T_c = 0$
because $\beta_c = \infty$. In Sec.~\ref{sec:1d} we present an
``extended scaling'' approach tailored to systems having $T_c =
0$ and a unique nearest-neighbor interaction strength $|J_{ij}|$
(in this case $|J_{ij}| = 1$ $\forall$ $i$, $j$ and no bond
disorder).\cite{comment:sg} We first present simple exact expressions
for the 1D Ising ferromagnet for which the scaling
variable
\begin{equation}
\tau_{\rm t}(\beta) = 1- \tanh(\beta)
\label{eq:taut}
\end{equation}
works well.  This is consistent with the Cardy {\em et al.}
rule\cite{cardy:01} because $\tau_{\rm t}$ is equal to $2\exp(-2\beta)$
at low temperatures; but, like $(1-\beta/\beta_c)$, $\tau_{\rm t}$
tends to $1$ for $T \to \infty$.  In the light of this result we then
apply the same approach using $\tau_{\rm t}$ to the nontrivial 2D
fully frustrated Villain model.  Our analysis shows that the extended
scaling scenario with $\tau_{\rm t}$ as a scaling variable gives
an excellent account of the behavior of the correlation length and
reduced susceptibility extrapolated to infinite size over the entire
temperature range.

\section{One-dimensional Ising model}
\label{sec:1d}

To motivate the scaling expressions for ferromagnetic Ising models with
zero transition temperature, we use as a toy model the one-dimensional
Ising ferromagnet
\begin{equation}
{\mathcal H}_{\rm 1D} = -\sum_{i = 1}^{L} J_{i, i+1}S_i S_{i+1}
\end{equation}
with $J_{i, i+1} = 1$ for all nearest neighbors $i$ and $i+1$.  The model
orders only\cite{baxter:82} at $T = 0$ and expressions for $\xi(\beta)$
and $\chi(\beta)$ in the infinite-size limit are easily calculated
from HTSE. The reduced susceptibility can be expanded as
\begin{equation}
\chi(\beta) =1+2[\tanh(\beta) + \tanh^2(\beta) + \tanh^3(\beta) + \ldots] 
\end{equation}
and the second moment of the correlation is
\begin{equation}
\mu_2(\beta) = 2[\tanh(\beta) + 2^2\tanh^2(\beta) + 3^2\tanh^3(\beta)
+ \ldots] .
\end{equation}
The second-moment correlation length is then given by Eq.~(\ref{eq:xi})
with $z = 2$ in 1D. Using the mathematical identities
\begin{equation}
\sum_{n=1}^{\infty} x^n = \frac{x}{1-x} \;\;\;\;\;{\rm and}\;\;\;\;\;
\sum_{n=1}^{\infty} n^2x^n = \frac{(x+1)x}{(1-x)^3} ,
\end{equation}
the exact expressions for susceptibility and correlation length are
thus
\begin{equation}
\chi(\beta)  = \exp(2\beta) = \frac{2}{1-\tanh(\beta)} - 1
\label{eq:chi1d}
\end{equation}
and
\begin{eqnarray}
\xi(\beta) &=& \frac{1}{2}[\exp(4\beta)-1]^{1/2} \nonumber \\
	   &=& \frac{\tanh^{1/2}(\beta)}{1 - \tanh(\beta)} \; .
\label{eq:xi1d}
\end{eqnarray}
Note that these expressions are valid for the {\em entire} temperature
range.

\begin{figure}
\includegraphics[width=0.95\columnwidth]{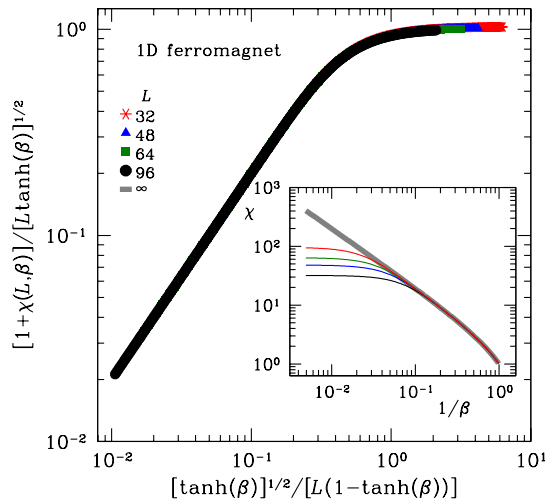}
\vspace*{-1.0cm}
\caption{(Color online)
Scaled ferromagnetic reduced susceptibility [Eq.~(\ref{eq:chi})]
for the 1D Ising ferromagnet according to Eq.~(\ref{eq:fsschi1d}).
The data scale perfectly and thus validate the derived scaling
expressions. The inset shows the unscaled data for different system
sizes, as well as the thermodynamic limit (thick gray line).  }
\label{fig:chiscale-fm-1d}
\end{figure}

Equations (\ref{eq:chi1d}) and (\ref{eq:xi1d}) are of the extended
scaling form\cite{campbell:08} [see Eqs.~(\ref{eq:chiext}) and
(\ref{eq:xiext})] with $\tau_{\rm t}$ [Eq.~(\ref{eq:taut})] replacing
$\tau$ in the extended scaling expressions for the ferromagnets with
finite ordering temperatures. Finally, temperature-dependent effective
exponents can be defined as
\begin{equation}
\gamma(\beta) = -d\log[\chi]/d\log[1 - \tanh(\beta)]
\label{eq:gammaTc0}
\end{equation}
and
\begin{equation}
\nu(\beta) = - d\log[\xi(\beta)/\tanh^{1/2}(\beta)]/d\log
[1 - \tanh(\beta)] \; .
\label{eq:nuTc0}
\end{equation}
In the limit $T \to T_c = 0$ the critical exponents are thus $\gamma_c
= \nu_c = 1$.

For the Ising ferromagnet in one space dimension with linear extent
$L = N$ the Fisher finite-size scaling rule\cite{yeomans:92} for
an observable
\begin{equation}
{\mathcal O}(L,\beta) \sim L^{y/\nu}{\mathcal F}[L/\xi(\beta)]
\label{FSS.1d}
\end{equation}
when applied to the reduced susceptibility leads to
\begin{eqnarray}
\frac{\chi(L,\beta) + 1}{L/\tanh(\beta)^{1/2}}
&\sim& {\mathcal F}_{\chi}
\left[ \frac{L[1-\tanh(\beta)]}{\tanh^{1/2}(\beta)} \right] \nonumber \\
&\equiv& {\mathcal F}'_{\chi}\left[\frac{\xi(\beta)}{L}\right] .
\label{eq:fsschi1d}
\end{eqnarray}
In Fig.~\ref{fig:chiscale-fm-1d} we illustrate the previously-derived
scaling relations with data for the reduced susceptibility for {\em
finite} system sizes.  The data are obtained by starting with the
partition function $Z  = \lambda_+^L+\lambda_-^L$ for a one-dimensional
system of $L$ spins in a field $H$,\cite{reichl:98} with
\begin{equation}
\!\!\!\! \lambda_\pm \! = 
\! e^{\beta}\!\!\left[\cosh(\beta H) \pm \sqrt{\cosh^2(\beta H) - 2
e^{-2\beta} \sinh(2\beta)}\right] .
\end{equation}
To obtain the thermodynamic susceptibility $\chi_{\rm th}(\beta)$,
we perform a second-order derivative of the free energy per spin,
$F=-(1/\beta)\ln Z(L)$, with respect to $H$, subsequently setting $H=0$.
The raw data for $\chi=\chi_{\rm th}/\beta$ (inset) are scaled
according to Eq.~(\ref{eq:fsschi1d}).  The scaling is perfect.

\section{Two-dimensional Villain model}
\label{sec:2d}

The two-dimensional fully frustrated Ising model, or Villain
model,\cite{villain:77} consists of Ising spins on a square lattice
with nearest-neighbor bonds $|J_{ij}| = 1$; in the $x$ direction all
bonds are ferromagnetic, while in the $y$ direction columns of bonds
are alternately ferromagnetic and antiferromagnetic. The Hamiltonian
is thus given by
\begin{equation}
{\mathcal H} = - \sum_{\langle i,j\rangle} J_{ij} S_i S_j \; ,
\label{eq:ham}
\end{equation}
where $S_i = \pm 1$ represent Ising spins on a square lattice with $N =
L^2$ spins. The system is fully frustrated; i.e., the product of the
bonds around {\em each} plaquette of the system is negative,
\begin{equation}
\prod_{\Box} J_{ij} = -1 \; .
\label{eq:ff_prod}
\end{equation}
The model does not order at a finite temperature,\cite{forgacs:80}
but exhibits a critical point at zero temperature with diverging
ferromagnetic reduced susceptibility and a ground-state degeneracy
which grows exponentially with the system size.

For the scaling analysis we compute the reduced susceptibility
[Eq.~(\ref{eq:chi})] and the finite-size second-moment
correlation length [Eq.~(\ref{eq:xiL})].  The simulations
are done using exchange (parallel tempering) Monte
Carlo\cite{geyer:91,hukushima:96,marinari:96} and the simulation
parameters are presented in Table \ref{tab:simparams}. Equilibration is
tested by a logarithmic binning of the data. Once the last three bins
for all observables agree within error bars the system is considered
to be in thermal equilibrium. We use periodic boundary conditions to
reduce finite-size corrections.

\begin{table}
\caption{
Parameters of the simulations. $L$ denotes the system size, $N_{\rm
sa}$ is the number of independent runs to improve the statistics and
$N_{\rm sw}$ is the total number of Monte Carlo sweeps performed
in a single run for each of the $2 N_T$ replicas.  $T_{\rm min}$
and $T_{\rm max}$ are the lowest and highest temperatures simulated,
respectively, and $N_T$ is the number of temperatures used in the 
parallel tempering method. \label{tab:simparams}
}
\begin{tabular*}{\columnwidth}{@{\extracolsep{\fill}} r r r r r r }
  \hline
  \hline
  $L$  &  $N_{\rm sa}$  & $N_{\rm sw}$ & $N_T$ & $T_{\rm min}$ & $T_{\rm max}$\\
  \hline
  $8$  & $1000$ &  $131072$ & $30$ & $0.1$ & $3.0$\\
  $12$ & $1000$ &  $131072$ & $30$ & $0.1$ & $3.0$\\
  $16$ & $1000$ &  $131072$ & $30$ & $0.1$ & $3.0$\\
  $24$ & $1000$ &  $262144$ & $30$ & $0.1$ & $3.0$\\
  $32$ & $1000$ &  $262144$ & $30$ & $0.1$ & $3.0$\\
  $48$ &  $500$ & $2097152$ & $30$ & $0.1$ & $3.0$\\
  $64$ &  $100$ & $2097152$ & $30$ & $0.1$ & $3.0$\\
  $96$ &  $100$ & $2097152$ & $30$ & $0.1$ & $3.0$\\
  \hline
  \hline
\end{tabular*}
\end{table}

Forgacs\cite{forgacs:80} showed analytically that the limiting
low-temperature thermodynamic behavior of the correlation length of
the 2D fully frustrated Villain model is strictly exponential, i.e.,
$\xi(\beta) \sim \exp(2\beta)$. Furthermore, the critical exponent
$\eta$ describing the decay of the correlation at $T_c$
is exactly $1/2$ such that in the low-temperature limit, using 
$\chi(\beta) \sim \xi(\beta)^{\gamma/\nu}$ 
[Eqs.\ (\ref{eq:chiext}) and (\ref{eq:xiext})] and the
standard scaling relation $\gamma=(2-\eta)\nu$, we obtain
$\chi(\beta)  \sim \xi(\beta)^{2-\eta} = \exp(3\beta)$. 
Based on an analysis of the size
dependence of the energy by Lukic {\em et al.}\cite{lukic:06} it has
been conjectured that the low-temperature limit for the correlation
length is exactly $\xi(\beta) = (1/2)\exp(2\beta)$. No full HTSE study
seems to have been carried out to date; however, by inspection, the
leading HTSE terms for the reduced susceptibility are $\chi(\beta) =
1 + 2\beta + O(\beta^2)$ and for the second moment of the correlation
length $\mu_{2} = \beta + O(\beta^2)$.

\begin{figure}
\includegraphics[width=0.95\columnwidth]{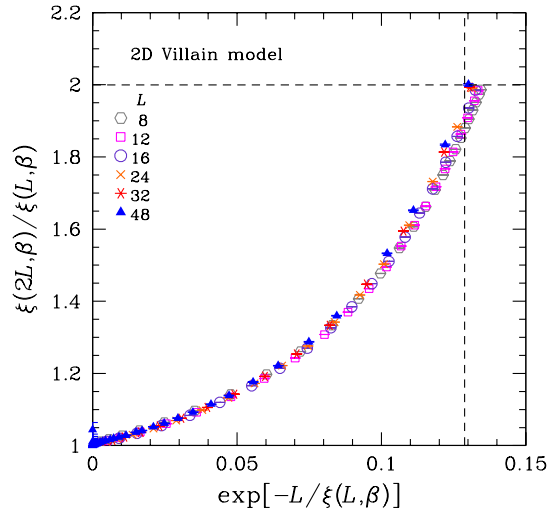}
\vspace*{-1.0cm}
\caption{(Color online)
Scaling ratio $\xi(2L,\beta)/\xi(L,\beta)$ for the 2D Villain model as
a function of $\exp[-L/\xi(L,\beta)]$ for different system sizes $L$.
The dashed horizontal line corresponds to the exact infinite size
value $\xi(2L,\beta_c)/\xi(L,\beta_c)= 2$ at the critical point.
The vertical line corresponds to the estimated infinite-size limit
$\xi(\beta_c)/L = 0.488$ (see text).
}
\label{fig:xi2LxiL-ff}
\end{figure}

\begin{figure}
\includegraphics[width=0.95\columnwidth]{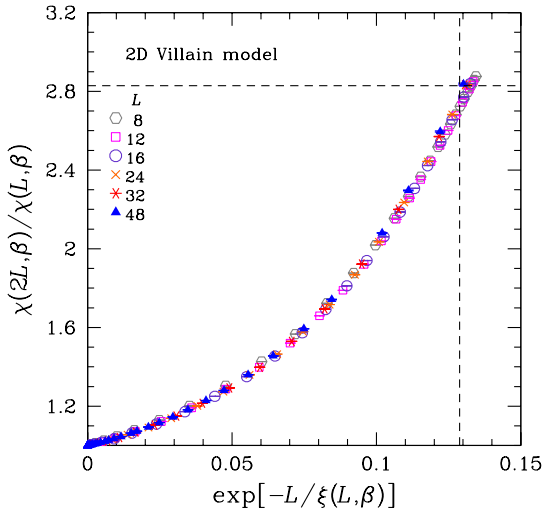}
\vspace*{-1.0cm}
\caption{(Color online)
Scaling ratio $\chi(2L,\beta)/\chi(L,\beta)$ as a function of
$\exp[-L/\xi(L,\beta)]$ for the 2D Villain model for different system
sizes $L$.  The dashed horizontal line corresponds to the exact
infinite-size value $\chi(2L,\beta_c)/\chi(L,\beta_c) = 2^{3/2}$ at
the critical point since in general $\chi(T_c,L) \sim L^{2 - \eta}$
and $\eta = 1/2$.  The vertical line corresponds to the estimated
infinite-size limit $\xi(\beta_c)/L = 0.488$ (see text).
}
\label{fig:chi2LchiL-ff}
\end{figure}

\begin{figure}
\includegraphics[width=0.95\columnwidth]{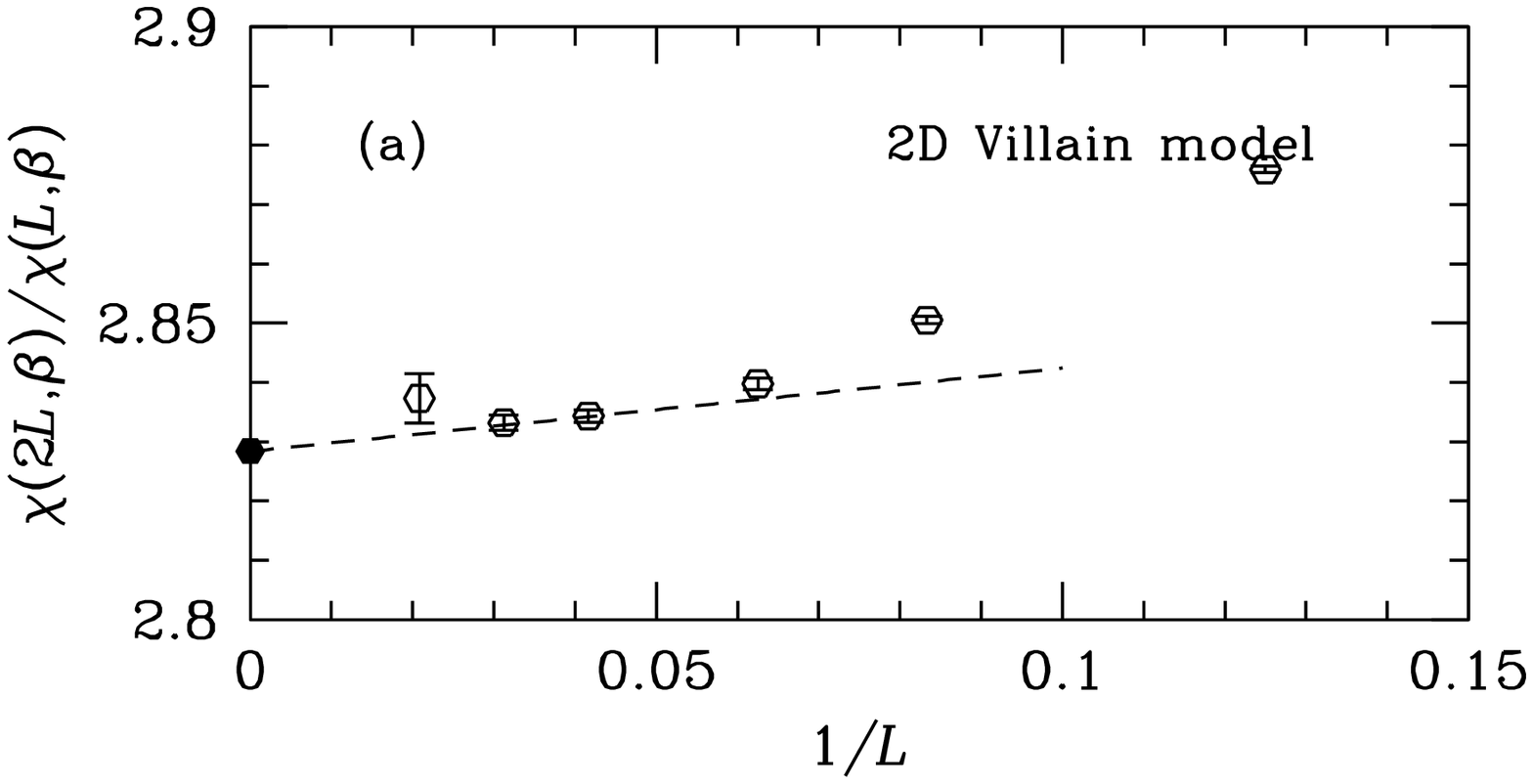}

\vspace*{-0.5cm}

\includegraphics[width=0.95\columnwidth]{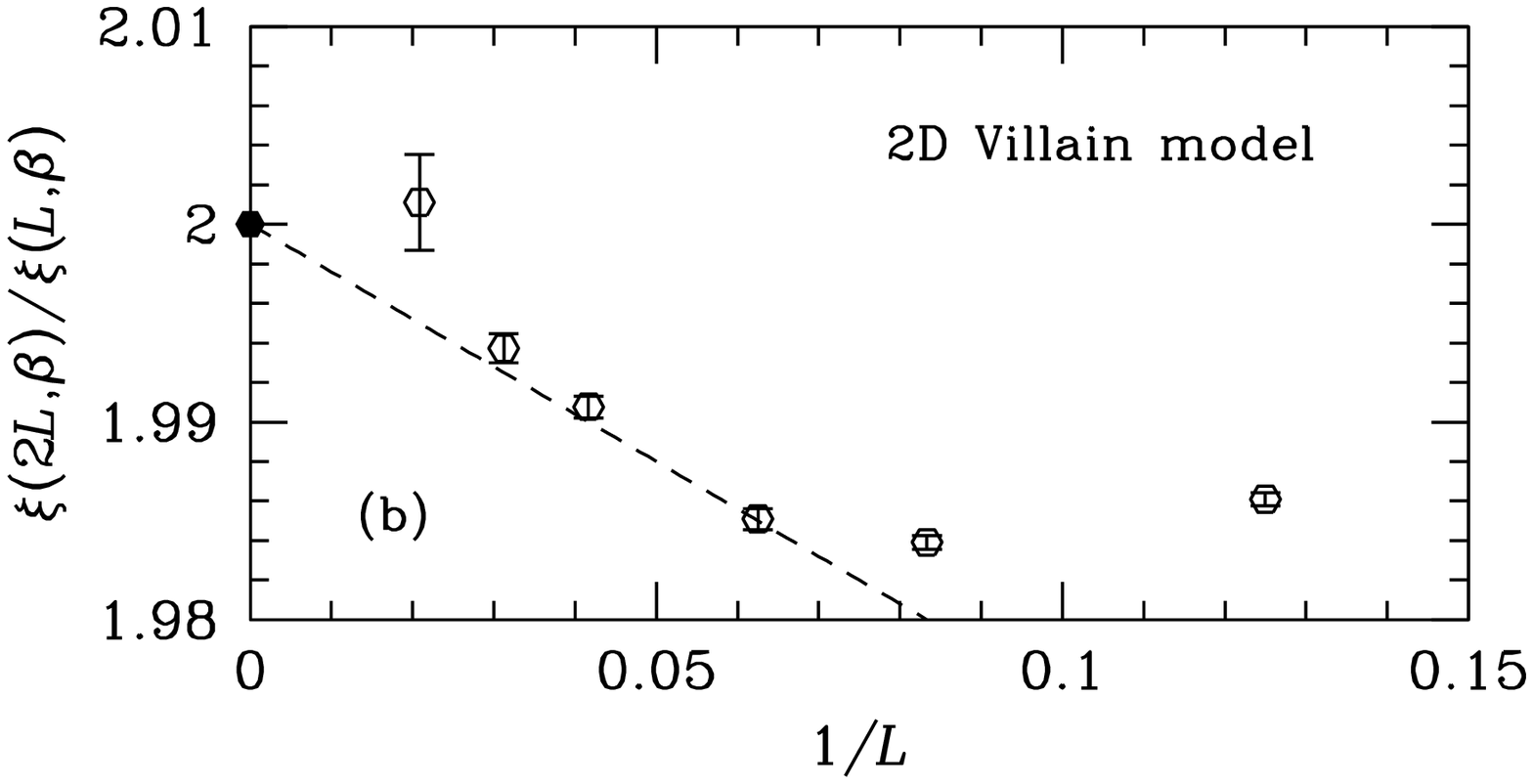}

\vspace*{-0.5cm}

\includegraphics[width=0.95\columnwidth]{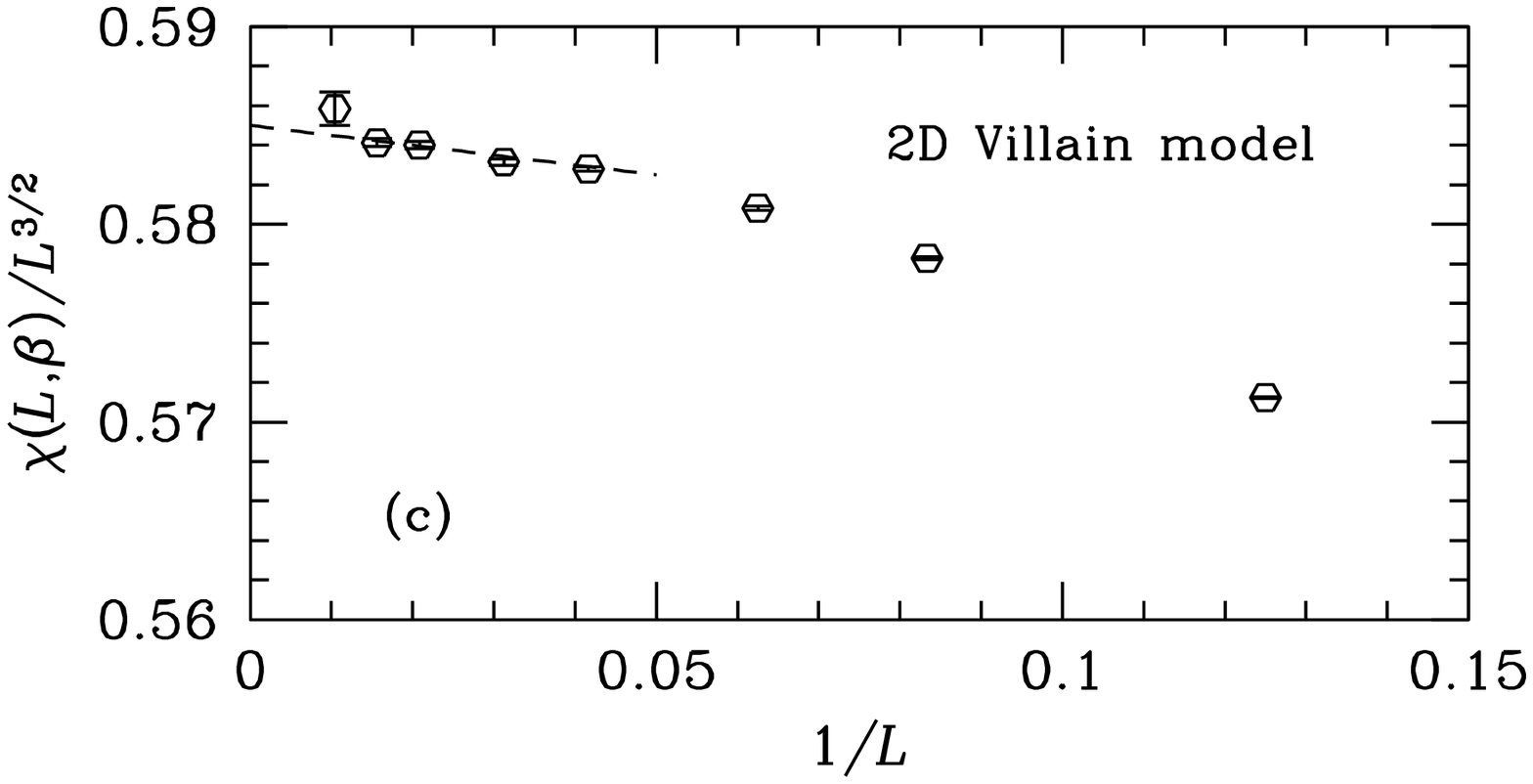}

\vspace*{-0.5cm}

\includegraphics[width=0.95\columnwidth]{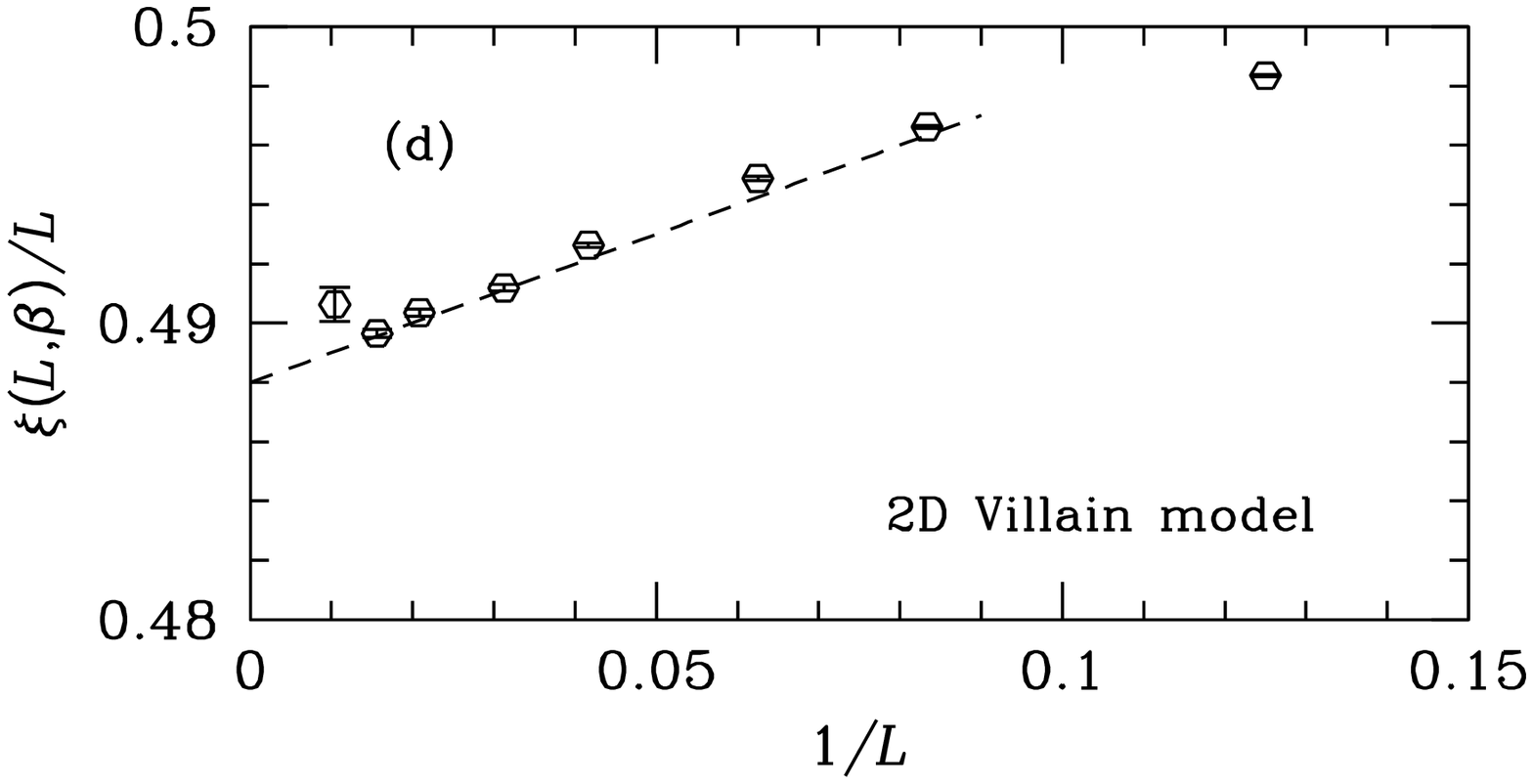}

\vspace*{-0.4cm}

\caption{(Color online)
Scaling ratios $\chi(2L,\beta)/\chi(L,\beta)$ [panel (a)] and
$\xi(2L,\beta)/\xi(L,\beta)$ [panel (b)] plotted against $1/L$. The
dashed line corresponds to $\chi(2L,\beta)/\chi(L,\beta) \sim 2^{3/2}
+ 0.14/L$ ($\xi(2L,\beta)/\xi(L,\beta) \sim 2 - 0.24/L$). The full
symbol corresponds to the exact thermodynamic value $2^{3/2}$ in
panel (a) and $2$ in panel (b). Panel (c): $\chi(L,\beta)/L^{3/2}$
vs $1/L$. The dashed line corresponds to $\chi(L,\beta)/L^{3/2}
\sim 0.585 - 0.05/L$.  Deviations appear for smaller values of $L$.
Panel (d): $\xi(L,\beta)/L$ plotted against $1/L$. The dashed line
corresponds to $\xi(L,\beta)/L \sim 0.488 + 0.1/L$. The data thus
suggest that a corrections to scaling exponent $\omega \approx 1$
is plausible.  All data are for $T = 0.10$. Note that the data point
for $L = 96$ is generally a bit high possibly due to the
small statistics used in the simulation.
}
\label{fig:corrections}
\end{figure}

Scaling dimensionless ratios of finite-size data for the
correlation length [$\xi(2L,\beta)/\xi(L,\beta)$] and susceptibility
[$\chi(2L,\beta)/\chi(L,\beta)$] vs  the two-point finite-size
correlation length divided by the system size [$\xi(L,\beta)/L$],
which is also a dimensionless quantity,\cite{caracciolo:95} yields
unique curves depending only on the universality class if there are no
finite-size corrections to scaling. For the system sizes studied, the
Villain model shows weak corrections to scaling. This can be seen in
Figs.~\ref{fig:xi2LxiL-ff} and \ref{fig:chi2LchiL-ff} where the ratios
are shown as functions of $\exp[-L/\xi(L,\beta)]$.\cite{caracciolo:95}
In principle, it should be possible to use the ansatz of Calabrese
{\em et al.};\cite{calabrese:03} i.e.,
\begin{equation}
\frac{\xi(2L,\beta)}{\xi(L,\beta)} = {\mathcal F}[L/\xi(L,\beta)]
			     + L^{-\omega}{\mathcal G}[L/\xi(L,\beta)] ,
\label{eq:cara}
\end{equation}
where $\omega$ is the finite-size scaling correction exponent and
${\mathcal F}$ and ${\mathcal G}$ are scaling functions. (Similar
relations apply for the susceptibility $\chi$.)  As shown below when $T
\to 0$, $\xi_{\infty}(\beta)$ diverges until $\xi_{\infty}(\beta)\gg
L$; the values of the observables saturate at $\xi(L,\beta) \to
\xi(L,\beta_c = \infty)$ and $\chi(L,\beta) \to \chi(L,\beta_c)$. The
lowest temperature at which the simulations have been carried out is
$T = 0.1$. At this temperature we find $\xi(L = \infty,\beta) \sim
2.8 \cdot10^{8}$, thus for all $L$ studied $\xi(L = \infty,\beta)
\gg L$ and we can take the measured values of observables at
all $L$ as good approximations to the $T = 0$ value.  Hence, for
$\beta\to\beta_c$ the pre-factor $\sin (|{\bf k}_\mathrm{m}|/2)$
is the only $L$-dependent factor in Eq.~(\ref{eq:xiL}), which
leads to $\xi(2L,\beta_c)/\xi(L,\beta_c) \to 2$. Furthermore,
because $\eta=1/2$, $\chi(2L,\beta_c)/\chi(L,\beta_c) \to
2^{2-\eta}= 2.82843 \ldots$ exactly.  Figure \ref{fig:corrections}(a) 
shows $\chi(2L,\beta_c)/\chi(L,\beta_c)$ and Fig.~\ref{fig:corrections}(b) 
shows $\xi(2L,\beta_c)/\xi(L,\beta_c)$ against $1/L$; it can be seen that
the ratio behaves approximately as $\chi(2L,\beta_c)/\chi(L,\beta_c)
\sim 2.8284 + 0.14/L$ [$\xi(2L,\beta_c)/\xi(L,\beta_c) \sim 2 -
0.24/L$] showing that the correction exponent for the leading
correction at large $L$ can be plausibly taken as $\omega \sim
1$ with further terms appearing at smaller $L$. In panel (c) of
Fig.~\ref{fig:corrections} we show data for $\chi(L,\beta_c)/L^{2-\eta}
= \chi(L,\beta_c)/L^{3/2}$ against $1/L$. Fitting the data assuming
$\omega = 1$ gives the large-size limit $\chi(L,\beta_c)/L^{3/2}
\approx 0.585(1) - 0.05/L$.  In a similar way we find the approximate
limiting value of $\xi(L,\beta_c)/L \approx 0.488(1) + 0.1/L$ [panel
(d)] and of the Binder cumulant $g(L,\beta_c) \sim 0.691(1)+0.10/L$
(not shown).

The analysis of the data at other temperatures is also consistent
with a leading correction with an exponent $\omega \sim 1$ plus
further correction terms for smaller $L$. At all temperatures studied
the difference between the estimated infinite-size values for the
observables and the measured large-$L$ values are always less than
$0.5\%$ of the measured large-$L$ values.

Inspired by the results on the 1D Ising ferromagnet outlined in
Sec.~\ref{sec:1d} with $\tau_{\rm t} = 1 - \tanh(\beta)$ as a scaling
variable we now test an analogous scaling of the data for the 2D
fully frustrated Ising model.

\begin{figure}
\includegraphics[width=0.95\columnwidth]{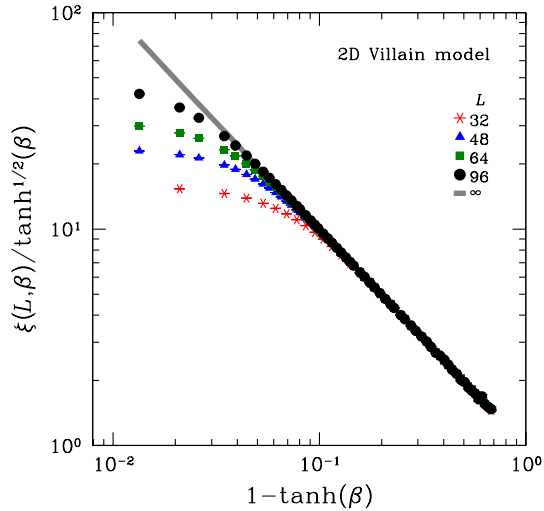}
\vspace*{-1.0cm}
\caption{(Color online)
Normalized correlation length $\xi(\beta)/\tanh^{1/2}(\beta)$
for the 2D fully frustrated Ising model for different system sizes
$L$ as a function of $1- \tanh(\beta)$. The thick line corresponds to
the extended scaling correlation length expression Eq.~(\ref{eq:xiFF}).
}
\label{fig:xiren-vs-tanh}
\end{figure}

The critical exponents for the 2D fully frustrated Ising model
[Eqs.~(\ref{eq:gammaTc0}) and ~(\ref{eq:nuTc0})] are  $\gamma_c = 3/2$,
$\nu_c = 1$, and $\eta_c = 1/2$.\cite{forgacs:80} We thus construct
trial expressions for the different observables as follows:
\begin{equation}
\chi_{\rm FF}(\beta) = C_{\chi}[1 - \tanh(\beta)]^{-3/2} + (1-C_{\chi})
\label{eq:chiFFtrial}
\end{equation}
and
\begin{equation}
\xi_{\rm FF}(\beta) = \tanh^{1/2}(\beta)\left[
\frac{C_{\xi}}{1-\tanh(\beta)} + (1-C_{\xi})\right] ,
\label{eq:xiFFtrial}
\end{equation}
where the critical amplitudes $C_{\chi}$ and $C_{\xi}$ are the only
adjustable parameters.  It turns out that for the correlation length,
the expression with $C_{\xi}=1.00(1)$, i.e.,
\begin{equation}
\xi_{\rm FF}(\beta)/\tanh^{1/2}(\beta) = [1 - \tanh(\beta)]^{-1}
\label{eq:xiFF}
\end{equation}
gives an excellent overall fit to
$\xi_{\infty}(\beta)/\tanh^{1/2}(\beta)$, which is the normalized
infinite-size limiting curve estimated from the scaling curves; see
Fig.~\ref{fig:xiren-vs-tanh}. Over the entire temperature range
the maximum difference between the fit and the numerical curve is
approximately $0.5\%$. The expression in Eq.~(\ref{eq:xiFF}) with
$C_{\xi}=1$ is identical to the exact expression for the 1D Ising
ferromagnet. In the low-temperature limit with $C_{\xi}=1$, $\xi_{\rm
FF} \to (1/2)\exp(2\beta)$ meaning that the present data and analysis
are consistent with the Lukic {\it et al.} conjecture\cite{lukic:06}
within numerical precision. In the high-temperature limit $\xi_{\rm
FF} \to \beta^{1/2}$, which is consistent with the first term of the
HTSE for $\xi(\beta)$.

\begin{figure}
\includegraphics[width=0.95\columnwidth]{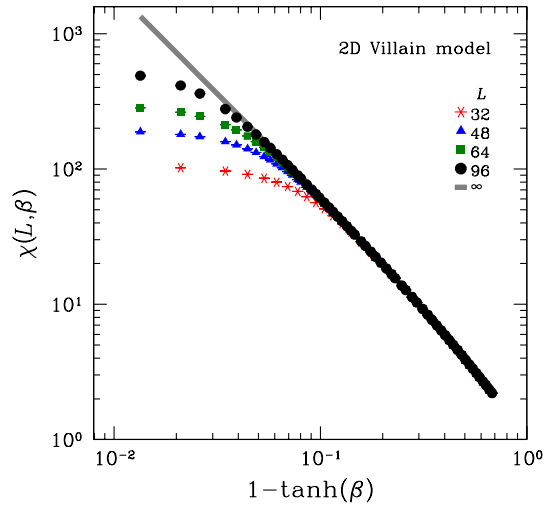}
\vspace*{-1.0cm}
\caption{(Color online)
Data for the susceptibility $\chi(L,\beta)$ of the 2D fully frustrated
Ising model for different system sizes $L$ as a function of $1-
\tanh(\beta)$. The thick line corresponds to the extended scaling
expression in Eq.~(\ref{eq:chiFF}).  }
\label{fig:chi-vs-tanh}
\end{figure}

For the reduced susceptibility the fits to the numerical data for
$\chi_{\infty}(\beta)$ with Eq.~(\ref{eq:chiFFtrial}) indicate that
$C_{\chi}$ is equal to $2.00(5)$; see Fig.~\ref{fig:chi-vs-tanh}. The
fit in the higher-temperature range can be improved further by a
correction term chosen so that there is an exact agreement between the
high-temperature limit obtained from the first two terms in the HTSE,
namely, $\chi(\beta) = [1 + 2\beta^2 + \ldots]$ as $\beta \to 0$.
We thus obtain:
\begin{equation}
\chi_{\rm FF}(\beta) = 2.0[1-\tanh(\beta)]^{-3/2} - 2.0 + [1- \tanh(\beta)] .
\label{eq:chiFF}
\end{equation}

\begin{figure}
\includegraphics[width=0.95\columnwidth]{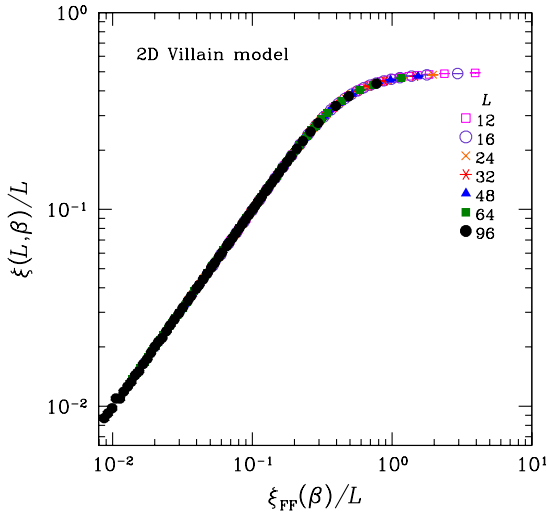}
\vspace*{-1.0cm}
\caption{(Color online)
Finite-size scaling of the two-point correlation length data of the
2D fully frustrated Ising model using the extended scaling expression,
Eqs.~(\ref{eq:xiFF}) and (\ref{eq:fssxi}).}
\label{fig:xi-vs-xiFF}
\end{figure}

\begin{figure}
\includegraphics[width=0.95\columnwidth]{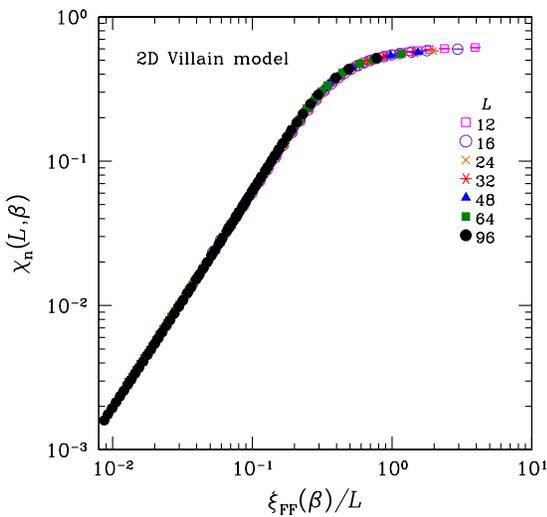}
\vspace*{-1.0cm}
\caption{(Color online)
Finite-size scaling of the susceptibility of the 2D
fully frustrated Ising model using the extended scaling expression
for the normalized susceptibility $\chi_{\rm n}(L,\beta)$,
Eqs.~(\ref{eq:chiFF}) and (\ref{eq:fsschi}).}
\label{fig:chi-vs-xiFF}
\end{figure}

With the extended scaling expressions given above
[Eqs.~(\ref{eq:xiFF}) and (\ref{eq:chiFF})] the standard Fisher finite-size 
scaling (FSS)
[Eq.~(\ref{FSS.1d})] is modified (see Refs.~\onlinecite{campbell:06}
and \onlinecite{campbell:07} for details). For the finite-size
correlation length we thus obtain from Eqs.~(\ref{FSS.1d}) and
(\ref{eq:xiFF})
\begin{equation}
\xi(L,\beta)/L \sim {\mathcal F}_{\xi}\left[
\frac{L[1-\tanh(\beta)]}{\tanh^{1/2}(\beta)}
\right] \equiv {\mathcal F}'_{\xi}\left[\frac{\xi_{\rm FF}}{L}\right] ,
\label{eq:fssxi}
\end{equation}
whereas for the normalized reduced susceptibility we obtain from
Eqs.~(\ref{FSS.1d}), (\ref{eq:chiFF}) and (\ref{eq:xiFF})
\begin{eqnarray}
\!\!\!\!\!\!\!\!\chi_{\rm n}(L,\beta) &\equiv&
\frac{[\chi(L,\beta) + 2 -(1-\tanh(\beta))]}{[L/\tanh(\beta)^{1/2}]^{3/2}} 
\nonumber\\
&\sim& {\mathcal F}_{\chi}
\left[ \frac{L[1-\tanh(\beta)]}{\tanh^{1/2}(\beta)} \right] 
\equiv {\mathcal F}'_{\chi}\left[\frac{\xi_{\rm FF}}{L}\right]  .
\label{eq:fsschi}
\end{eqnarray}
A finite-size scaling analysis of the data for the second-moment
correlation length and the susceptibility using Eqs.~(\ref{eq:fssxi})
and (\ref{eq:fsschi}) is shown in Figs.~\ref{fig:xi-vs-xiFF} and
\ref{fig:chi-vs-xiFF}, respectively.

The scaling curves have a characteristic form. Quite generally, at
small $\xi_{\infty}/L$, $\xi(L,\beta)/L = \xi(L = \infty,\beta)/L$
so the log-log plot of, e.g., Fig.~\ref{fig:xi-vs-xiFF} is initially a
straight line of slope $1$ passing through the point $[1,1]$. At the
large $\xi_{\infty}/L$ limit (which is equivalent to $T = 0$ for all
$L$) the curves tend to plateau values, $K_{\chi} = \chi(L)/L^{3/2} =
0.585(1)$ and $K_{\xi} = \xi(L)/L = 0.488(1)$ estimated above. If we
ignore the marginal corrections to finite-size scaling, the crossover
can be expressed phenomenologically as
\begin{equation}
\xi(L,\beta)/L =
\left[
\frac{[\xi_{\infty}(\beta)/L]^{z_{\xi}}}{1
+(1/K_{\xi})^{z_{\xi}}[\xi_{\infty}(\beta)/L]^{z_{\xi}}}
\right]^{1/{z_{\xi}}} ,
\end{equation}
where $z_{\xi}$ is a crossover exponent.
In the present case $z_{\xi} \approx 2.5$.  For the
reduced susceptibility the initial small-$\xi_{\infty}/L$ behavior
is $\chi(L,\beta)/L^{2-\eta} \sim [\xi_{\infty}(\beta)/L]^{(2-\eta)}$
and the analogous phenomenological crossover equation is
\begin{equation}
\chi(L,\beta)/L^{2-\eta} =
\Lambda
\left[\frac{[\xi_{\infty}(\beta)/L]^{z_{\chi}(2-\eta)}}{1
+(\Lambda/K_{\chi})^{z_{\chi}}[\xi_{\infty}(\beta)/L]^{z_{\chi}(2-\eta)}}
\right]^{1/z_{\chi}}  ,
\end{equation}
where $z_{\chi}$ is the crossover exponent, $\Lambda$ is a constant,
and $K_{\chi}$ is the plateau value. The phenomenological fit values
in the present case are $z_{\chi} \approx 2.0$ and $\Lambda \approx
1.95$.
For both observables the fits
with crossover are of excellent quality.

\section{Summary and Conclusion}

We have presented scaling expressions motivated via high-temperature
series expansions which extend the scaling functions across the whole
temperature range for systems which order at zero temperature.

For the 1D Ising ferromagnet we can derive exact extended scaling
expressions of the form $\chi(\beta) = 2\tau_{\rm t}^{-1} - 1$ and
$\xi(\beta) = \tau_{\rm t}^{-1}\tanh^{1/2}(\beta)$ with the scaling
variable $\tau_{\rm t} = 1 - \tanh(\beta)$, critical exponents
$\gamma_{c} = \nu_{c} = 1$ [defined via Eqs.~(\ref{eq:gammaTc0}) and
(\ref{eq:nuTc0})], and critical amplitudes $C_{\chi}= 2$ and $C_{\xi}= 1$.

From the insights gained from the study of the 1D ferromagnet, we
use the same temperature variable $\tau_{\rm t}$ to analyze numerical
data for the 2D fully frustrated (Villain) model.  The exact critical
exponents are known:\cite{forgacs:80} $\gamma_{c}= 3/2$ and $\nu_{c}=
1$. We find that for the second-moment correlation length $\xi(\beta)
= \tau_{\rm t}^{-1}\tanh^{1/2}(\beta)$ with $C_{\xi} = 1$ just as
for the 1D ferromagnet.  Furthermore, this result is consistent within
numerical accuracy with the low-temperature-limit conjecture of Lukic
{\it et al.}\cite{lukic:06} that $\xi(\beta) \to (1/2)\exp(2\beta)$
for $T \to 0$; however the present expression covers the {\em entire}
temperature range.  The approximate expression for the susceptibility
of the Villain model [Eq.~(\ref{eq:chiFF})] with critical amplitude
$C_{\chi}=2.00(5)$ is in good agreement with the numerical data over
the entire temperature range covered.

Summarizing, the temperature dependence of observables above
a ferromagnetic transition (including $T_c = 0$ transitions)
can be written in terms of generic extended scaling
forms\cite{campbell:06,campbell:07,campbell:08} expressed to leading
order as
\begin{equation}
\xi(x) = x^{-1/2}[C_{\xi}(1-x)^{-\nu} +(1-C_{\xi})]
\label{eq:xigen}
\end{equation}
and
\begin{equation}
\chi(x) = C_{\chi}(1-x)^{-\gamma} + (1-C_{\chi})
\label{eq:chigen}
\end{equation}
with the scaling variable $x$ and critical parameters
depending on the system studied. The expressions are exact by
construction in the limits $\beta \rightarrow \beta_c$ and $\beta
\rightarrow 0$ if the critical parameters are known.  For a
ferromagnet with $T_c > 0$ $x= \beta/\beta_c$. Note that in the
high-dimensional limit Eqs.~(\ref{eq:xigen}) and (\ref{eq:chigen}) are
exact.\cite{campbell:08} In finite dimensions (but with nonzero $T_c$)
the expressions remain as good approximations over the entire temperature
range. For the two ferromagnets with $|J_{ij}| = 1$, $T_c = 0$, and
no bond disorder, $x = \tanh(\beta)$ and thus $\tau_{\rm t} = [1 -
\tanh(\beta)]$ replaces $[1 -\beta/\beta_c]$ as the scaling variable.
Effective exponents are defined through Eqs.~(\ref{eq:gammaTc0}) and
(\ref{eq:nuTc0}). These relations are validated with numerical data
on the 2D fully frustrated Ising model.

There are numerous possible candidate systems to which this approach
should in principle be applicable {\em mutatis mutandis}. These
include for instance the family of fully frustrated 2D systems
studied by Forgacs,\cite{forgacs:80} the 2D three-state Potts
antiferromagnet,\cite{caracciolo:95,cardy:01}, the 2D Ising
antiferromagnet on a triangular lattice,\cite{wannier:50} the
2D $\sigma$ models,\cite{caracciolo:93,caracciolo:95} as well
as 2D Heisenberg models.\cite{caracciolo:95a} An interesting
further step would be to determine scaling expressions for the
two-dimensional bimodal Ising spin glass with $|J_{ij}| = 1$ but
with random signs for the interactions, which also orders at zero
temperature.  In that case the critical behavior of the model is highly
controversial\cite{houdayer:04,lukic:04,katzgraber:05b,joerg:06a,katzgraber:07c,hartmann:08}
and current data at finite temperature do not have the necessary
quality within ``traditional'' scaling approaches to determine the
true nature of the transition.

\begin{acknowledgments}

We would like to thank Alan Sokal for helpful comments.
The simulations were performed on the ETH Z\"urich computer
clusters. H.G.K.~acknowledges support from the Swiss National Science
Foundation under Grant No.~PP002-114713 and would like to thank the
Aspen Center for Physics for their hospitality.

\end{acknowledgments}

\bibliography{refs,comments}

\begin{thebibliography}{33}
\bibitem[{com({\natexlab{a}})}]{comment:escale}
\bibinfo{note}{For all discussed models we set the absolute value of the
  spin-spin interactions $|J_{ij}| = 1$ $\forall$ $i$,$j$.}
\bibitem[{com({\natexlab{b}})}]{comment:sg}
\bibinfo{note}{The method can also be applied to spin glasses at zero
  temperature provided that $\tanh(\beta)$ is replaced by $\tanh^2(\beta)$.
  Unfortunately, current numerical data do not have the necessary quality to
  cleanly test the extended scaling on this model. Note that random-field
  ferromagnets ordering at zero temperature can be described with the
  zero-temperature extended scaling relations.}
\end{thebibliography}

\end{document}